# OAM-Assisted Self-Healing Is Directional, Proportional and Persistent

**Marek Klemes, Lan Hu, Greg Bowles, Alireza Ghayekhloo, Mohammad Akbari, Soulideth Thirakoune, Michael Schwartzman, Kevin Zhang, Tan Huy Ho, David Wessel and Wen Tong**
Canada Research Center, Huawei Technologies Canada Co., Ltd., Kanata, Ontario, CANADA.
Corresponding author: (e-mail: marek.klemes@huawei.com).

This work was supported in part by the Canada Research Center, Huawei Technologies Canada Co., Ltd., Kanata, Ontario, CANADA.

**ABSTRACT** In this paper we demonstrate the postulated mechanism of self-healing specifically due to orbital-angular-momentum (OAM) in radio vortex beams having equal beam-widths. In previous work we experimentally demonstrated self-healing effects in OAM beams at 28 GHz and postulated a theoretical mechanism to account for them. In this work we further characterize the OAM self-healing mechanism theoretically and confirm those characteristics with systematic and controlled experimental measurements on a 28 GHz outdoor link. Specifically, we find that the OAM self-healing mechanism is an additional self-healing mechanism in structured electromagnetic beams which is directional with respect to the displacement of an obstruction relative to the beam axis. We also confirm our previous findings that the amount of OAM self-healing is proportional to the OAM order, and additionally find that it persists beyond the focusing region into the far field. As such, OAM-assisted self-healing brings an advantage over other so-called non-diffracting beams both in terms of the minimum distance for onset of self-healing and the amount of self-healing obtainable. We relate our findings by extending theoretical models in the literature and develop a unifying electromagnetic analysis to account for self-healing of OAM-bearing non-diffracting beams more rigorously.

**INDEX TERMS** Self-healing beams, orbital angular momentum (OAM), millimeter waves, Helmholtz equation, non-diffracting beams, Bessel beams, vortex beams, helical phase, helical Poynting vector, partial beam obstruction, structured light, focusing region, far field, Butler matrix, DFT matrix, transverse wave vectors, axial wave vector.

## I. INTRODUCTION

As the density of wireless connections continues to grow with each generation of wireless technology, it is inevitable that an increasing proportion of those connections will encounter physical obstructions which will necessitate employment of non-line-of-sight (NLOS) transmission techniques. Indeed, the diversity of spatial paths resulting from NLOS propagation offers new degrees of freedom for multiplexing streams of data signals which have been exploited by multi-input-multi-output (MIMO) techniques involving multi-element antenna arrays at both ends of mobile and fixed wireless links.

However, as the demand for wireless data transmission capacity continues to grow, the technology is migrating into higher frequencies of the electromagnetic spectrum where more bandwidth is available, but the radio wavelengths are correspondingly smaller, i.e. millimeter-waves (mm-waves). Consequently, path losses also increase, and must be offset by higher antenna gains to maintain adequate signal-to-noise ratios (SNR). This leads to narrower beam-widths which illuminate fewer scatterers so mm-wave MIMO systems resort to large antenna arrays to generate multiple spatially-diverse beams to achieve their degrees of freedom. The remaining scatterers are also less penetrable so they become obstructions at these wavelengths.

Nevertheless, even these multiple narrow beams will inevitably encounter more obstructions as their numbers continue to grow with increasing demand for wireless data capacity in a finite physical environment. Inevitably fewer obstruction-free sites will be available for base-station antennas and a greater proportion of mobile users will have no clear line of sight (LOS) to them.

Consequently, alternative techniques of spatial multiplexing are being developed for mm-waves and higher radio frequencies. One approach uses Reconfigurable Intelligent Surfaces (RIS) which re-direct the narrow beams toward users under intelligent control. This involves additional wireless infrastructure and signaling overhead. Another approach is borrowed from the optical community, where use is made of structured light beams, specifically non-diffracting beams carrying Orbital Angular Momentum (OAM) [1]. Such beams possess particular profiles of phase and amplitude in the transverse plane orthogonal to their axis, which gives them the ability to re-focus and re-form their structures after being partially obstructed.

In this work we are specifically interested in radio-frequency (RF) OAM beams which possess these self-healing properties, whereby power flow from different parts of their transverse structures fills in portions of the beam that have been removed by obstructions. Because the receiving antenna in an RF wireless link never captures the entire transmitted beam, the self-healing process is not required to be perfect as long as it can restore the portion of the beam seen by the receiving antenna system.

It turns out that uniform circular arrays (UCA) are particularly well suited to generating OAM beams. Exciting the elements with linear phase progressions resulting in integer multiples of $2\pi$ radians covering the UCA circumference gives the OAM beams a hollow conical shape in their far fields and a helical phase progression around their common axis, as depicted in Figure 1. It also endows the OAM beams with corresponding orthogonal-helical power flow which may be exploited for self-healing. Such phase progressions with uniform amplitudes are readily obtainable from a Discrete Fourier Transform (DFT) matrix. Indeed, RF beams having helical phase around their axis have already been synthesized for other applications decades ago [2] but were utilized in the azimuth plane orthogonal to their axis. They were otherwise known as phase modes, and were readily synthesized using circular arrays whose elements were fed from a modified Butler matrix, which was in fact equivalent to the radix 2 form of the Fast Fourier Transform (FFT) matrix.

Other structured beams can also possess self-healing properties in their non-diffracting regions [3, 5, 8], but here we are interested in the role of OAM in the self-healing of mm-wave beams, which is not specifically well covered in the literature. We wish to gain insight into how it would allow for siting OAM antennas where partial obstructions due to vegetation or people may occur, and for countering obstruction losses in the far-field in potential 6G applications.

In previous work we have characterized a 28 GHz millimeter-wave wireless OAM downlink in terms of spatial multiplexing [4]. More recently we have observed self-healing properties of the OAM beams that we employed [6]. In this work we describe additional experiments with OAM beams targeted to characterize and validate the self-healing mechanism postulated in [6]. Furthermore, we relate the self-healing mechanism to self-healing mechanisms of other structured beams found in the literature and combine them into a unified OAM-assisted self-healing mechanism of structured beams possessing OAM, which shows that OAM enhances their self-healing properties. We specifically show that our experimental results demonstrate that the OAM-assisted self-healing effect is ***proportional*** to the OAM order, ***directional*** with respect to the OAM beam axis according to the sign (polarity) of the OAM order, and that it ***persists*** into the far field, beyond the Rayleigh range of the OAM beams.

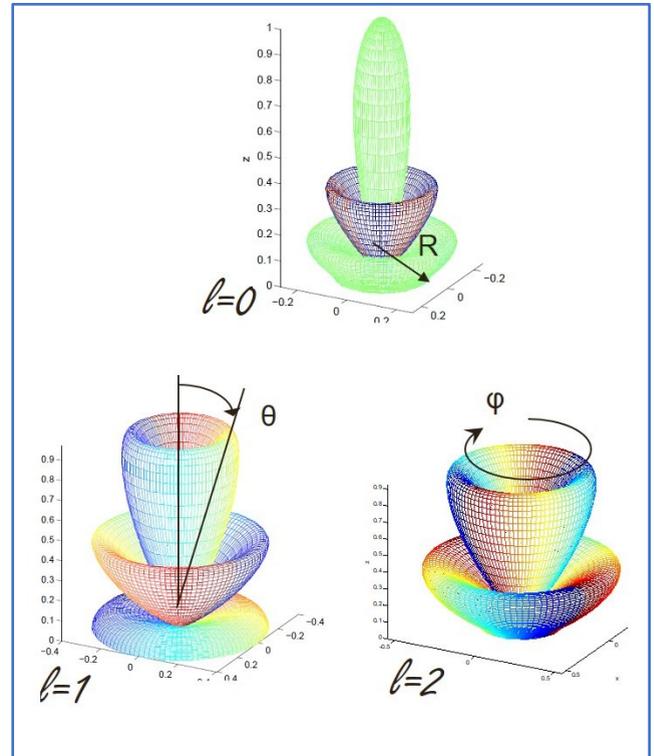

**Figure 1.** Examples of OAM beams of order *l* = 0, +1, +2 generated by a uniform circular array. Colour denotes electrical phase, amplitude is normalized to peak value and propagation is along z axis (vertical).

These objectives were motivated by an observation about our previous results where a physical mechanism for OAM-assisted self-healing was proposed [6]. There it was developed from the guide wavelength in a coaxial waveguide and a phase-continuity condition in its transverse direction, similar to the boundary conditions of a rectangular waveguide for the fundamental $TE_{10}$ mode. The open end of the postulated coaxial waveguide of mean radius *R* was treated as the source aperture of our OAM antenna and the radius of our presumably non-diffracting OAM beams. It was then determined in equation (12) of [6] that guide wave number $k_z=2\pi/\lambda_z$ and free-space wave number $k=2\pi/\lambda$ form a right-angled triangle relationship with $k_{T\ell}=\ell/R$, which was therefore also treated as a wave vector, but in the circumferential ($\psi$ direction) in the transverse plane orthogonal to the *z* axis of the beam. (A radially-directed transversal wave vector component present in structured non-diffracting beams was ignored in [6] for clarity.) This $\psi$-directed transverse wave vector was deemed responsible for the self-healing effects observed, but it was left for future work to characterize the OAM self-healing mechanism more rigorously.

Accordingly, upon a physics-based analysis of self-healing mechanisms found in the literature [9, 10], it was noted that theoretical expressions for ***all*** of the electromagnetic components of structured beams carrying OAM mode of order $\ell$ and propagating in the axial $z$ direction, contain the complex-valued phase factors

$$e^{j\omega t} e^{-j(l\varphi + k_z z)}$$

It is customary to omit the omnipresent time-dependent factor, where $\omega = 2\pi f$ with $f$ being the RF carrier frequency in Hz. Here $k_z = 2\pi/\lambda_z$ with $\lambda_z$ being the axial-direction wavelength, and $\psi$ is the circumferential angle in the transverse plane of the beam. It therefore seemed natural to write the OAM phase factor as a product of a wave vector and propagation distance just like the phase factor due to propagation in the $z$ direction. The wave vector in this case would be the postulated one for OAM order $\ell$, $k_{T\ell} = \ell/R$, and the corresponding propagation distance would be the arc length around the circumference of the beam at radius $R$ given by $\psi R$, so that

$$l\varphi = (l/R)(\varphi R) = k_{Tl}(\varphi R) \qquad (1)$$

and

$$e^{-j(l\varphi + k_z z)} = e^{-j(k_{Tl}(\varphi R) + k_z z)} \qquad (2)$$

This appears to justify the postulated OAM-assisted mechanism as propagation of power in the transverse direction, comparable to the usual propagation of power in the beam's axial direction along the $z$-axis. Additionally, the orbital angular momentum per photon can be derived from this formulation similarly as linear momentum per photon is derived using the free-space wave vector (which is also the direction of power flow given by the Poynting vector ***S***). Thus it can be established that the self-healing observed in OAM beams is indeed partly due to their orbital angular momentum, which causes their power flow direction to form a spiral Poynting vector orthogonal to the spirals of equi-phase surfaces. Furthermore, this OAM-assisted self-healing is ***directional*** in accordance with the sign of $\ell$, ***proportional*** according to the magnitude of $\ell$, and ***persists*** all the way from the source into the far field as $z \to \infty$ because the corresponding phase factor in (2) is always present in the electromagnetic wave equation for OAM beams. It is precisely these aspects that the present experiments and analyses are intended to confirm.

This paper is organized as follows: Section II describes and partial-obstruction strategy for demonstrating the directionality, proportionality and persistence properties of OAM-assisted self-healing, the experimental set-up of millimeter-wave OAM antenna and link, and the measurement strategy targeting both the hypothesized OAM-enabled self-healing properties and the "control" measurements using beams of the same angular width but not containing OAM. Section III presents the key results of the experiments and relates them to the properties of the hypothesized self-healing mechanism. Section IV further confirms the OAM-assisted mechanism supported by our experiments by references to similar self-healing mechanisms analyzed in the literature and extends these to show how OAM enhances the self-healing ability of structured beams which possess OAM. Section V draws the conclusions confirming our hypothesized OAM-assisted self-healing mechanism and summarizes its enhanced properties relative to those of similar-sized non-OAM beams. Appendix A derives the complex amplitude describing a non-diffracting OAM beam from Maxwell's equations for propagating fields based on reference [10]. It also formulates a new wave vector in the transverse plane, thereby unifying the relations among the various wave vectors and momentum components responsible for the non-diffracting properties described in the literature and the additional self-healing properties of OAM beams as revealed in this study.

Insights gained from our investigations are also highlighted, with implications for propagation modelling and site planning of wireless networks and prospects for increasing their spectral efficiency as future MIMO techniques expand into the mm-wave spectrum and beyond.

## II. EXPERIMENTAL mm-WAVE OAM OUTDOOR LINK, OBSTRUCTION AND MEASUREMENT STRATEGY

The OAM experimental test bench used in this work was designed as a one-way LOS link with the multi-mode mm-wave OAM antenna transmitting an independent data stream on each of up to 5 mutually-orthogonal OAM beams to one multi-channel user equipment (UE) fitted with a linear array of 4 antennas and receivers (RX). More details are contained in references [4, 6] and itemized in Table 1. The transmitting (TX) base-station (BS) OAM antenna in particular is detailed in references [6, 7].

Layout of the link showing transmitters with OAM antenna, movable absorber panels serving as the obstruction, and 4-element receiving (RX) antenna array with a 4-channel UE are shown in Figure 2. The RF was selected as 28 GHz, for a typical link length of 100m, with a far field beginning at 30m. The obstruction was made of RF absorber panels so as to eliminate complicated interference patterns that would be caused by re-radiation from currents induced on a metallic obstruction. The panels were secured to movable partitions on wheels to assure repeatability. To minimize interference from multipath, right-hand circular polarization (RHCP) was employed at both ends to reject ground reflections by ~ 12dB, as all the equipment was mounted on movable carts at less than 2m above ground. The TX antenna was tilted up by half of the cone angle of the beams so as to position the RX at the bottom arc of all the conical OAM beams.

**Table 1. Experimental OAM link and TX antenna parameters**

| Parameter | Symbol | Relation | Value | Units |
|---|---|---|---|---|
| **Desired link specification.** | | | | |
| Bandwidth | $B$ | (LTE channel) | 20 | MHz |
| Number of OAM modes | $K$ | Co-channel data streams | 5 | 1 |
| Link distance | $L$ | TX to RX | 100 | m |
| RX antenna separation | $d$ | Max. UE antenna spacing | 0.20 | m |
| Digital IF of RX | $F$ | ADC bandwidth | 983.04 | MHz |
| **Dependent link parameters** | | | | |
| OAM beam radius | $R$ | $R < dF/(2B)$ $R = 0.87\text{m}$ | <2.4576 | m |
| RF wavelength | $\lambda$ | $28.0 \times 10^9 / 3.0 \times 10^8$ | 0.011 | m |
| TX antenna radius (max). | $r$ | $r \sim (K-1)\lambda L/(4\pi R)$ | 0.201 | m |
| TX far field | $L_{far}$ | $2(2r)^2/\lambda$ | 32.35 | m |
| No. of TX array elements (max) | $N$ | $N \sim 2\pi r/(\lambda/2)$ | 238 | 1 |

For our present purposes, OAM beams +2, -2, +4, -4 and non-OAM beam "0" were transmitted separately before inserting the obstruction, then again after inserting the obstruction. Additionally, the TX antenna was rotated in equal 4 steps of 11.25 degrees to average the measurements over the ripples in the OAM phase gradients seen at the RX, due to the discretized OAM excitation by 8 and 16 probes in the UCAs of the TX. The difference in signal levels and SNR caused by the obstruction was then noted for each UE receiver channel. These differences were then averaged over the 4 RX channels for each OAM mode, and tabulated for each position of the obstruction.

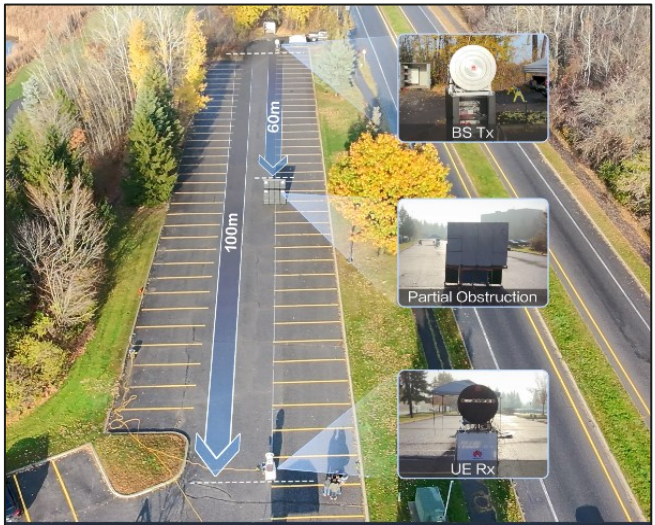

**Figure 2.** Drone's eye view of experimental outdoor OAM link and 28 GHz transmitting and receiving equipment, with movable absorber obstruction between them.

A separate set of experiments was also conducted using a narrower non-OAM beam from the same aperture as the OAM beams, to determine the effect of OAM order versus non-OAM on self-healing ability in the presence of an on-axis symmetric obstruction. Only positive OAM orders +2, +4 and 0 were utilized in those tests to avoid any asymmetries due to polarity and to obstruction placement relative to the beam axis.

***Most importantly, all of the OAM beams needed to have the same cone angles so that they would overlap at the receiving end***, because each beam carried a separate data stream to it on the same common RF carrier frequency, which would need to be de-multiplexed in UE. Additionally, it was important to exclude the effects of differing beam-widths from these experiments, as we were interested in the effects of only the OAM orders and their polarities. Therefore the TX antenna was composed of 3 concentrically stacked circular-array antenna discs, with the larger ones transmitting the correspondingly higher orders of OAM beams, OAM+4 and OAM-4, the middle disc for OAM+2 and OAM-2, and the smallest disc for the control beam, OAM_0. Its illustration is reproduced from our earlier work [6] in Figure 3, and the effective alignment of the cone angles of OAM modes ±4 with those of OAM±2 based on their theoretical Bessel-function beam shapes is reproduced in Figure 4.

Each UCA was a type of radial line slot array (RLSA), fed with a small circular array of probes around the center of a radial waveguide. Its far-field radiation pattern of the $\ell$-th OAM mode can be analytically approximated by coherently summing the radiated fields from the elements of a UCA resulting in the equation

$$F_\ell(\theta, \varphi) = (-j)^\ell e^{j\ell\varphi} K J_\ell\left(\frac{2\pi r \sin\theta}{\lambda}\right) \quad (3)$$

where $r$ is the radius of the transmitting circular (actually spiral) array radiating the $\ell$-th OAM mode, $\lambda$ is the radiating wavelength, $J_\ell$ is the $\ell$-th order Bessel function of the first kind, $K$ is a constant of propagation, $\theta$ is the radian half-cone angle or tilt angle from the beam ($z$) axis while $\varphi$ is the azimuth angle around the beam axis as well as an integer sub-multiple of the electrical phase of the beam at that azimuth angle. (The reason why the UCA's were spirals rather than pure circles was that the crossed-slot elements radiating RHCP hat to be aligned along the radii at each position, which caused their orientation to rotate in step with their angular position in the UCA. This had the effect of rotating the electrical phase of their RHCP emitted fields, which would add to the order of OAM radiated by the UCA. To compensate for that, the element excitations were delayed in phase in the radial direction by the same amount they were rotated in the azimuth direction, which had the effect of forming a spiral which advanced linearly along the radial direction until it was one wavelength rather out when it reached the starting point. This was deemed an insignificant departure from the ideal design and confirmed through detailed electromagnetic simulations.)

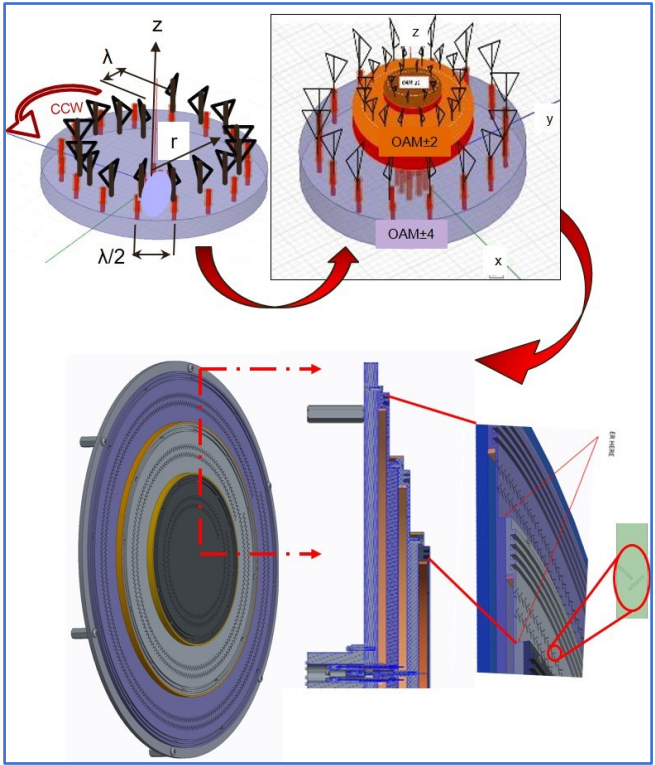

**Figure 3.** Design of multi-mode OAM antenna, showing radially-oriented RHCP array elements, arranged in a clockwise outward spiral to compensate rotation phase of RHCP, stacked in discs of increasing radii for increasing OAM orders, with cross-section detail and detail of RHCP crossed-slots array elements [6].

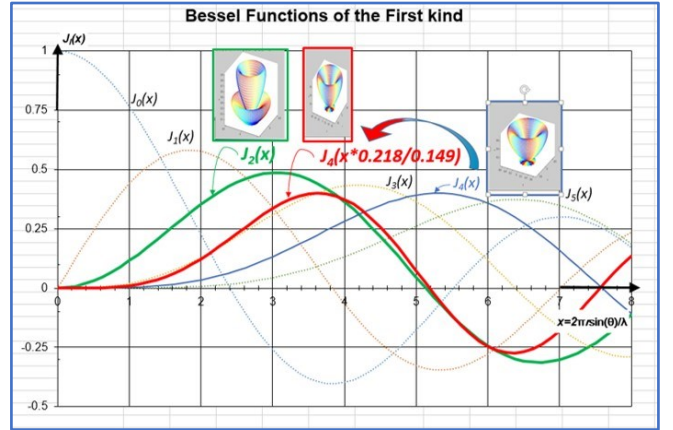

**Figure 4.** Scaling of OAM+4 amplitude profile to make its half-cone angle the same as half-cone angle of OAM+2 in the far field. OAM+2 is sourced from aperture of diameter 0.149m (green) so OAM+4 (blue) must be sourced from aperture of diameter 0.218m (red) to overlap with OAM+2 beam at the receiver [6].

The radial profiles of the OAM beams follow equation (1), each RLSA radius $r$ being designed so that the amplitude patterns of the higher and lower-order OAM beams coincide in the far field with $r_{OAM2} = 0.149$m and $r_{OAM4} = 0.218$m, as illustrated in Figure 4, where $x = k\, r \sin\theta \approx 3.5$ and $k = 2\pi/\lambda$. Note that the half-cone angle $\theta$ of the OAM±2 beams (green curve and green-outlined inset), and the scaled OAM±4 beams (red curve and red-outlined inset) are now both equal, corresponding to a half-cone angle of about 2.5 degrees from the beam axis. This scaling of the OAM amplitude profiles then leaves their *azimuth (or circumferential) phase profiles as the main structural difference between the OAM±2 and OAM±4 beams*.

The receiver array was placed, at $\theta \sim 2.5$ degrees down from the beam axis by tilting the TX antenna upward by the same angle. Transmitted signal levels were also adjusted to be equal at the receiver in all of the beams. Received signal levels were measured on time-orthogonal pilot signals appended to the spatially-multiplexed data signals. The pilot signal for each OAM beam was a pseudo-BPSK (pseudo-random symbols {(1+j), (-1-j)} at 30.72 Msymbols/s) signal sampled at 122.88 Msamples/s.

Having outlined the salient aspects of the general setup, we next describe the measurement strategy to test our hypotheses. Figure 5 captures graphically the test scenario with the hypothesized features. In each test case, the magnitude of self-healing effect in an individually-transmitted OAM beam was determined as the inverse of the difference in received signal level at each RX antenna caused by inserting the partial obstruction, in dB. (This separates out any effects of MIMO-type algorithms that combine the RX antenna signals in the UE.) In other words, smaller path loss caused by inserting the obstruction denoted greater amount of self-healing. Note that this does not imply self-healing of the entire beam, only the portion (arc) "seen" by the UE antenna array.

The main hypothesis we desired to test was that, if OAM contributes to self-healing, its effect should be directional, i.e. the amount of self-healing for a negative off-axis displacement of the obstruction should be greater (or smaller) in a negative-polarity OAM beam than in a positive-polarity OAM beam, and the reverse should be true for a positive off-axis displacement of the obstruction. The reasoning is that self-healing happens by power flow from other parts of the beam according to the hypothesized transverse wave vectors, $k_{T\ell} = \ell/R$, whose **directions** are given by the OAM polarities. As a control, any self-healing or other effects in a non-OAM beam (i.e. OAM_0) should show no preference with direction of obstruction displacement from the beam axis.

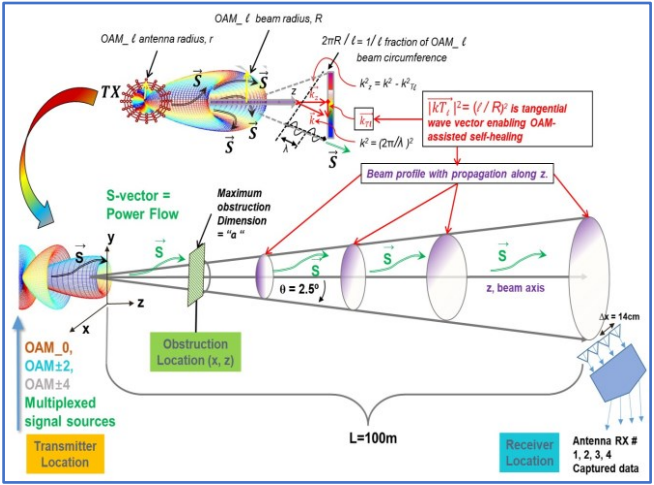

**Figure 5.** Test scenario for directionality, proportionality and persistence of OAM-assisted self-healing.

Next, we also hypothesized that self-healing effects in higher-order OAM beams would be greater in each case than in lower-order ones with the same cone angle, consistent with our observations in previous work [6]. There it was hypothesized that the magnitudes of the postulated transverse wave vectors responsible for the self-healing power flow were *proportional* to the magnitude of the self-healing effect. Additionally, self-healing effects in OAM beams should be greater than in non-OAM beams of similar angular width.

Finally we wished to ascertain that the OAM-assisted *self-healing effects persist* into the far field, as would be expected from the azimuth-dependent phase factor in any beam carrying OAM, because that factor is present even in the near field (where structured non-OAM also beams experience self-healing), as well as in the excitations and aperture source field distribution of OAM beams, as will be shown in Appendix A. This feature would make OAM beams more useful than traditional directional "pencil" beams in mm-wave wireless links, especially those using multiple beams as MIMO spatial degrees of freedom, because it would lead to lower (better) channel-matrix condition numbers and therefore greater spectral efficiencies.

## III. RESULTS AND INTERPRETATIONS OF MEASUREMENTS WITH PARTIALLY-OBSTRUCTED OAM BEAMS AND RELATIONS TO HYPOTHESES

In this section we present the results of the signal measurements captured by the UE receivers and processed off-line to show the effects of the partial beam obstructions. Figure 6 captures the processed experimental measurements that show basic agreement with all 3 of our hypotheses in annotated numerical form. Specifically, for every position of the obstruction on the z-axis, each of 2 successive rows corresponds to alternating displacements of the obstruction at +x and –x positions of its center from the z-axis. (At larger distances along the z-axis, the width of the obstruction was doubled to cover the correspondingly larger width of the beams, so the x-displacements of its center were also doubled from 60 cm to 120 cm.)

As for the columns of the table in Figure 6, for every order of OAM mode, each of 2 adjacent columns corresponds to $+\ell$ and $-\ell$, i.e. positive and a negative polarity of that order, with the right-most column corresponding to the control beam OAM_0.

| ALTERNATING OBSTRUCTION DISPLACEMENT DIRECTIONS AND OAM POLARITIES IN NEAR AND FAR FIELDS | Added Link Pathloss due to Partial Blockage with fixed UE Rx and blockage at various z-distances and alternating x-offsets from z-axis | | | | |
|---|---|---|---|---|---|
| Obstruction Positions (+x is right, -x is left looking from RX) | OAM+2 | OAM-2 | OAM+4 | OAM-4 | OAM_0 |
| Blockage at z=10m at x-offset = +60cm | -11.25 | -6.34 | -13.22 | -4.10 | -9.40 |
| Blockage at z=10m at x-offset = -60cm | -8.07 | -12.65 | -6.23 | -17.66 | -7.61 |
| Blockage at z=20m at x-offset = +60cm | -8.59 | -5.48 | -10.10 | -4.32 | -6.94 |
| Blockage at z=20m at x-offset = -60cm | -9.60 | -11.57 | -7.24 | -14.20 | -10.15 |
| Blockage at z=30m at x-offset = +60cm | -10.77 | -7.93 | -12.70 | -8.76 | -7.33 |
| Blockage at z=30m at x-offset = -60cm | -12.01 | -13.89 | -10.84 | -16.66 | -10.39 |
| Blockage at z=40m at x-offset = +120cm | -10.97 | -8.28 | -12.02 | -8.24 | -8.29 |
| Blockage at z=40m at x-offset = -120cm | -13.98 | -16.16 | -12.56 | -16.47 | -12.98 |
| Blockage at z=60m at x-offset = +120cm | -14.01 | -11.49 | -14.37 | -12.63 | -11.78 |
| Blockage at z=60m at x-offset = -120cm | -11.14 | -11.77 | -9.33 | -13.28 | -9.63 |

**Figure 6.** Dependence of self-healing in +OAM and –OAM polarities on alternating +x and –x obstruction displacements from beam axis, compared to independence of non-OAM beam.

It can readily be observed by comparing adjacent entries for each OAM mode in Figure 6 that the smaller obstruction losses highlighted in green always correspond to the same polarity of OAM mode and opposite sign of obstruction displacement. Specifically, this implies greater self-healing being achieved at +x displacements with OAM-$\ell$, and at –x displacements with OAM+$\ell$ consistently, while any self-healing of the OAM_0 control beam shows no preference with direction of x-displacement. The staggered pattern of highlighted cells attests to the consistency of this behavior, which confirms our *directionality* hypothesis.

It can also be noted in the same table, that the highlighted obstruction losses in each row are lower for higher-order modes OAM±4 than for the lower-order modes OAM±2. There are only 2 exceptions to this pattern, indicated by the red numbers. In other words higher-order OAM beams achieve greater self-healing than lower-order ones, for each OAM polarity and corresponding sign of obstruction displacement off-axis. The non-highlighted cells signify that there is essentially no self-healing taking place, so it serves no purpose to compare those in each row. This observation confirms our hypothesis that OAM-assisted self-healing is *proportional* to OAM order. In this case the entries for the OAM_0 control beam do not always show the highest obstruction losses as would be expected from the hypothesized pattern, because its shape was vastly different from the hollow conical OAM beams, its central main lobe being less obstructed especially at greater distances of the obstruction from the TX along the *z* axis. The green highlighted cells in the control column show where the self-healing pattern does agree with the proportionality hypothesis. (Taller obstructions would evidently be required to effect a comparable blockage of the OAM_0 control beam at the greater distances in the *z* direction from the TX.)

Finally we note from the same results in Figure 6 that the directionality and proportionality behaviours of OAM-assisted self-healing *persist* into the far field at distances of z > 30m from the TX. This is consistent with our persistence hypothesis about OAM-assisted self-healing, in contrast to non-OAM self-healing of other structured beams which is in effect only in their non-diffracting region of z, and directly determined by the transverse dimension of the aperture source of their fields. In fact the ideal non-diffracting beam has source amplitude distribution dependent on Bessel functions which have infinite support in the transverse direction, but because practical apertures are finite, the resulting pseudo-Bessel beams have only a finite non-diffracting range in the propagation direction [11]. The OAM property of such beams is typically pointed out in connection with higher-order Bessel beams [10, 11], but this aspect is rarely brought together with investigation of self-healing except in [12] and [9], where the helical Poynting vector is actually illustrated as being orthogonal to the equi-phase helical surfaces of an OAM beam.

Seeing that the phase helicity of an OAM beam is present throughout its propagation range, it is instructive to further examine its associated orthogonal-helicity of the Poynting vector which determines the direction of power flow in it. Being orthogonal to the helical equi-phase surface, the Poynting vector helix actually winds in the *opposite* direction to that of the phase helix as the beams progresses along its z-axis. When viewing the OAM beam as coming toward the viewer, the equi-phase helix of a positive-order OAM beam winds in the clockwise (CW) direction and that of a negative-order OAM beam winds in the counter-clockwise (CCW) direction, as the phase factor is taken to advance the phase of the beam with azimuth. Because the vectors representing power flow must be orthogonal to the respective equi-phase helices, it is easy to visualize that they will wind in the opposite directions and at smaller angles to the z-axis. These are the Poynting vectors denoted by ***S*** in Figure 5 which are responsible for filling in the blocked portions of an OAM beam as it propagates. The postulated transverse wave vector is responsible for "pushing" the power-flow ***S***-vector sideways along the conical main lobe of an OAM beams to form the helix. In the subsequent Figures, the ***S***-vector helices are indicated as viewed along the z-axis, annotated with the signs of their associated OAM modes.

In Figure 7 the obstruction at *z*=20m is displaced by 60cm in the –x direction and the power flow indicated by the curved arrows is CW for OAM+ℓ and CCW for OAM-ℓ. Those directions also correspond to the rotation directions of the shaded regions of the OAM beam "damaged" by the obstruction as denoted by the small smaller red circular beam profiles at the obstruction location (*z*=20m), whereas the large blue circles denote the beam profiles seen by the RX at *z*=100m (drawn to scale). That is to say the shaded damaged region rotates in the winding direction of the helical ***S***-vector of the corresponding OAM beam. It is therefore clear that the left-side damaged region of the OAM+ℓ beam rotates CW, away from the RX antennas at the bottom of the RX beam and brings more power from the undamaged part of the beam to them. This mechanism supports self-healing as denoted by the green highlighted entries in Figure 6.

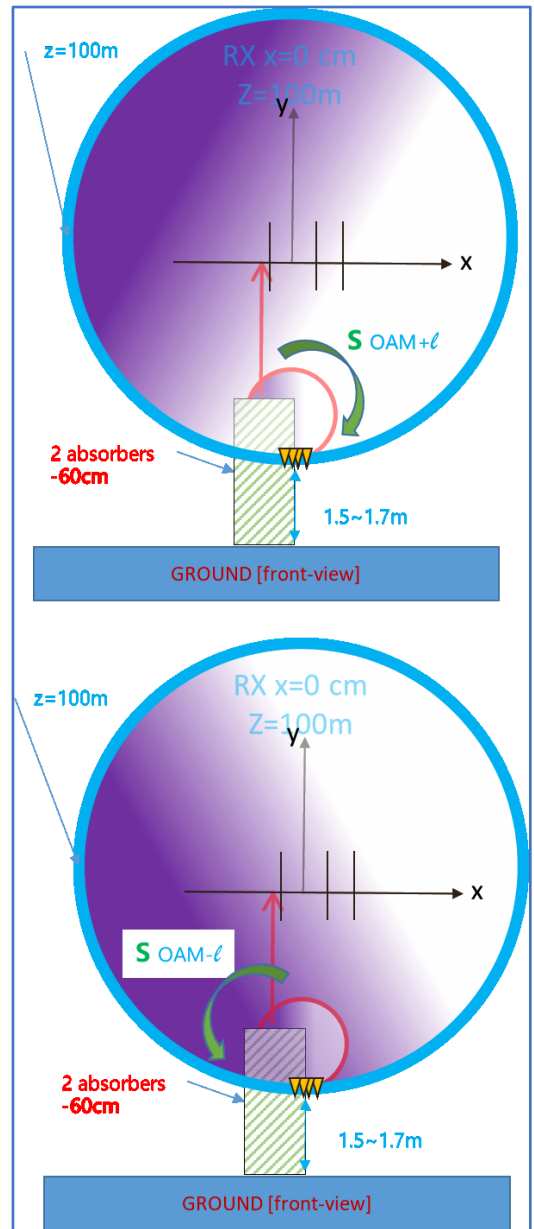

**Figure 7.** Obstruction at –x favours self-healing of + OAM beams.

Conversely, the CCW rotating ***S***-vector of the OAM-ℓ beams rotates the damaged region more toward the RX antennas at the bottom of the RX beam, thus not supporting any significant self-healing action for them.

Similarly, when the obstruction at *z*=20m is displaced in the +x direction as in Figure 8, we see the opposite effects: The damaged region of the bam is now on the right side (like the obstruction) so the ***S***-vector rotations favour healing the OAM-l beams and not the OAM-ℓ beams, as again evidenced in the table of Figure 6.

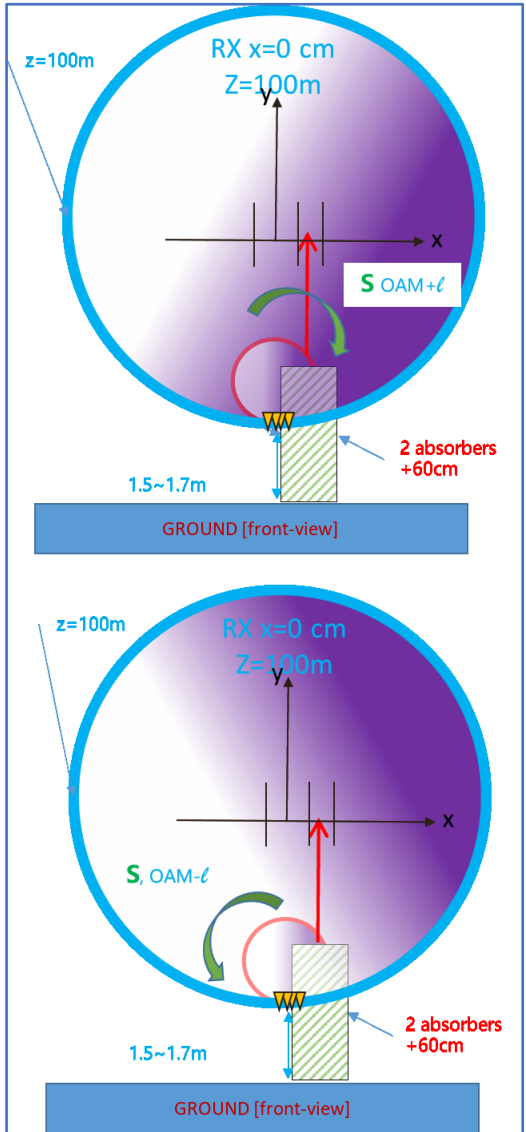

**Figure 8.** Obstruction at +x favours self-healing of − OAM beams.

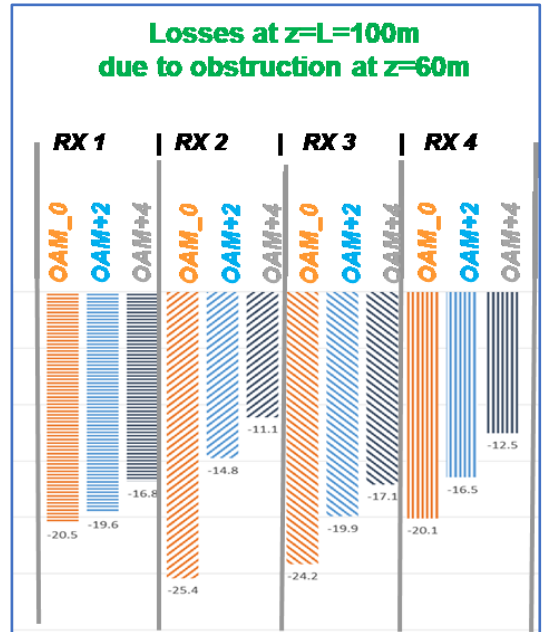

**Figure 9.** Effects of partial obstruction of beam OAM+4 on the correlations of pilot signals received on each RX antenna.

A separate set of experiments was performed with a narrower non-OAM beam sourced by the same aperture, and up-tilted by a smaller angle so as to illuminate the obstruction and receiver antennas approximately equally as the OAM beams. A sample of the results is presented in Figure 9 below, showing that the proportionality pattern of self-healing includes the non-OAM beam (OAM_0) and operates well into the far field.

The obstruction was placed on-axis at x = 0 cm and z = 60m and only positive OAM polarities were utilized to isolate the proportionality behavior from any directionality behaviour of the self-healing phenomena. It is clearly observed that the obstruction losses are smallest for the highest-order OAM beams and largest for the zero-order (non-OAM) beam at each of the 4 receiver antennas, meaning that the amount of self-healing was highest in the high-order OAM and lowest in the lowest-order OAM beams.

Although the directionality of self-healing in OAM beams is reported in the optical literature, it is not investigated systematically in terms of its relation to OAM polarity and order. The authors of [9,15] do investigate the skew angle of the Poynting vector which is linked to the azimuthal transverse wave vector and gives it its helicity, but they do not consider self-healing properties. On the other hand, the authors of [16] explain the self-healing mechanism of Bessel beams using Babinet's principle and wave optics, including higher-order Bessel beams (HOBB) carrying OAM, but they attribute the self-healing entirely to the radial transverse wave vector and make no mention of the azimuthal wave vector present in HOBBs.

Similar disjointedness can be found among other samples of the literature, so in the next section we develop a more unified model of self-healing in OAM beams and show how OAM improves their self-healing performance in line with the hypotheses as confirmed by our results above.

## IV. BENEFITS OF OAM-ASSISTED SELF-HEALING AND EXTENSIONS OF NON-DIFFRACTING SELF-HEALING MECHANISMS

Reference [12] serves as a convenient starting point for our geometric self-healing model using wave vectors. Although it relates to Bessel beams which are not exactly well emulated by our OAM antenna, the OAM properties still apply in the model. Bessel beams are a more rigorous formulation of non-diffracting beams. They will be treated in Appendix A to substantiate our model of OAM-assisted self-healing by relating it to particular solutions of Maxwell's equations, and the momentum relations derived from them. Reference [17] also describes self-healing mechanisms along the same lines as [12] and points out their relation to Bessel beams as the exact solutions to the Helmholtz wave equation derived from Maxwell's equations. The same argument was applied to substantiate the design of a truncated-Bessel beam antenna similar to ours in [11], so it is safe to assume it is applicable in our case also.

According to reference [17], a Bessel beam can be formed by illuminating a conical prism known as an axicon with a plane wave. Because the prism must have a finite diameter, $D$, it is a truncated Bessel beam of half-cone angle $\theta$ and no OAM, to begin with. (Incorporating a spiral phase plate into the axicon will endow the beam with OAM and make it a HOBB [17, Fig. 1 (e)].) The angle $\theta$ is determined by the prism angle $\gamma$ and refractive index, $n$, of the axicon according to the lens-maker's equation as

$$\theta = \sin^{-1}(n \sin \gamma) - \gamma \quad (4)$$

Its depth of focus, which is the non-diffraction region along its axis is then derived from the simple geometry of Figure 10 as

$$DOF = \left(\frac{D}{2}\right)\frac{1}{\tan \theta} \quad (5)$$

It may be argued that the cone half-angle $\theta$ in the far field is the same as angle $\theta$ here for simplicity, so we may use the development in [8] to find the minimum distance past an obstruction placed in the non-diffracting near-field region where self-healing begins.

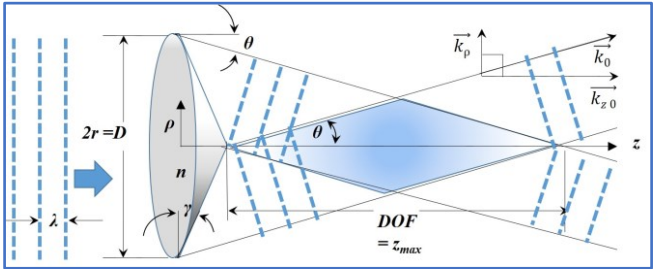

Figure 10. Geometry of generating a non-diffracting truncated Bessel beam of order 0 by illuminating an axicon prism with a plane wave [12].

Adapting the illustration of the self-healing distance from [17] and placing an obstruction of diameter $2a$ inside the depth of focus of Figure 10 above results in the situation depicted in Figure 11 below. The obstruction forms a shadow region to the right of it which has the length $z_{min}$ along the beam axis, after which the beam begins self-reconstructing. From the geometry it is readily determined that

$$z_{min} = \left(\frac{2a}{2}\right)\frac{1}{\tan \theta} = \frac{a}{\tan \theta} \quad (6)$$

which agrees with the general expression in reference [8]. Note that the same geometry of Figure 10 also applies in Figure 11.

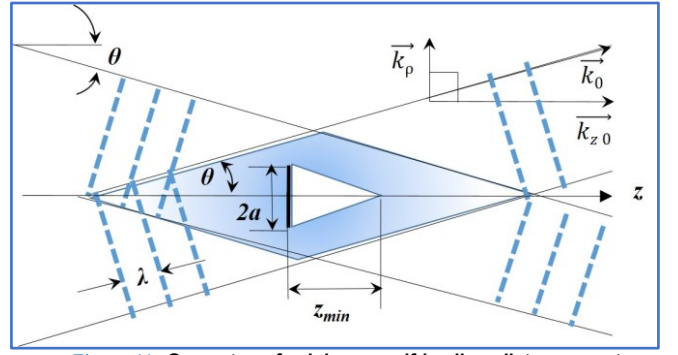

Figure 11. Geometry of minimum self-healing distance past an obstruction placed in the near-field focal region of a non-diffracting beam.

Note also from the geometry of Figures 10 and 11, that the radial and the orthogonal axial wave vectors add to form the free-space wave vector and their magnitudes, which are the wave numbers, are related by the Pythagorean theorem as

$$k_0^2 = k_z^2 + k_\rho^2 \quad (7)$$

and also $k_z = k_0 \cos\theta$ and $k_\rho = k_0 \cos$ so

$$\tan \theta = \frac{k_\rho}{k_z} \quad (8)$$

where $k_0 = 2\pi/\lambda_0$ and $k_z = 2\pi/\lambda_z$. This holds for non-diffracting beams which do not possess OAM, such as the truncated zero-order Bessel beam, hence the subscript "0". The same mechanism responsible for the non-diffracting property, namely the radial transverse wave vector, $k_\rho$, is also responsible for self-healing of non-diffractive beams.

Next we include OAM of order $\ell$ using the postulated tangential wave vector $k_{T,\ell} = \ell/R$ and note how it affects the self-healing performance. Note that this additional transversal wave vector is orthogonal to the radial transversal wave vector $k_\rho$ in the $\{\rho, \varphi\}$ plane, in the direction of the second of the cylindrical coordinates, $\varphi$, and so also obeys the Pythagorean relationship. It is therefore natural to include both transverse wave vectors as well as the axial wave vector to compose the complete 3-dimensional free-space wave vector which we will now call $k=2\pi/\lambda$, as

$$k^2 = k_z^2 + \left(k_\rho^2 + k_{T,\ell}^2\right) \quad (9)$$

The bracketed quantity in equation (9) can be considered as the transversal wave vector modified by the OAM tangential wave vector, which accounts for the skewing of the Poynting vector caused by orbital angular momentum of "$\ell$" per photon [15]. This skew angle can be visualized as the angle by which the transverse $\{\rho, \varphi\}$ plane is rotated around the vertical axis (indicated by $\rho$ in Figure10), but only for a particular value of $\varphi$. Figure 12 attempts to show the geometric relationship among the wave vectors. It holds for a particular value of OAM phase because OAM phase increases with azimuth angle $\varphi$, and the free-space wave vector **k**, which is the direction of the Poynting vector **S**=**E** ⨯**H**, must always be orthogonal to it, thus forming a helix as OAM phase increases. The helical shape of the Poynting vector is also explicitly shown also in [9, Figure 1], although no mention is made of self-healing.

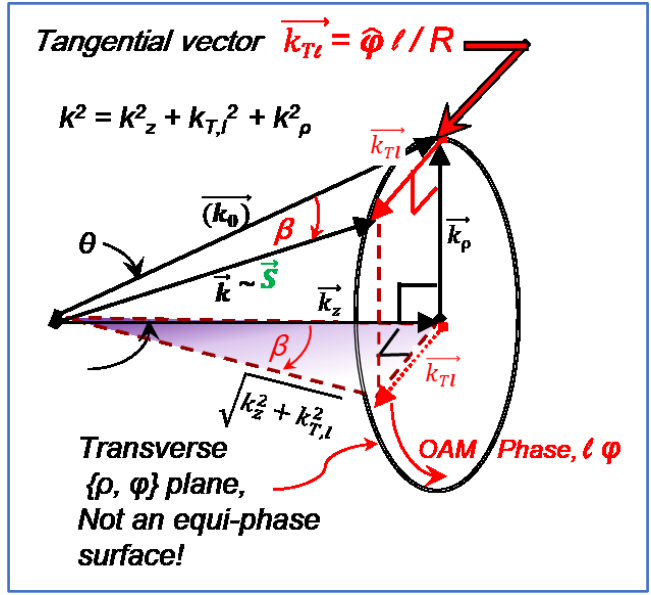

**Figure 12.** Geometry of minimum self-healing distance past an obstruction placed in the near-field focal region of a non-diffracting beam The features shown in red are due to OAM.

The skew angle of the Poynting vector (as well as the equi-phase surfaces) relative to the beam axis can be deduced from the geometry of Figure 12 as

$$\sin \beta = \frac{k_{T,l}}{\sqrt{k_z^2 + k_{T,l}^2}} \approx \beta \qquad (10)$$

which also agrees with [15] for small angles if $k_\rho$ is negligible in the far field beyond the DOF. Interestingly, reference [9] contains the phase factor we identified in the left side of equation (2) in every one of the electromagnetic field components of vector vortex waves. If our concept of the $\varphi$-directed transverse wave vector was adopted in [9], the skew angle relation (5) therein could be reformulated as

$$\beta = \sin^{-1}\left(\frac{k_{T,l}}{\sqrt{k_z^2 + k_{T,l}^2}}\right) \qquad (11)$$

with the correspondence that $\beta \rightarrow$ "Tilt Angle", $\theta \rightarrow \delta$, $R \rightarrow \rho$, $\ell \rightarrow m$, $k_{T,\ell} = \ell/R$, and $k_z \rightarrow (2\pi/\lambda)\cos\delta$, which agrees with (10).

Having thus established the correspondence of the published self-healing models of non-diffracting beams without OAM, and with skew angles of vortex beams carrying OAM with our OAM-assisted self-healing model via their various wave vectors, we proceed to extend the published results for self-healing distance and self-healing amounts by substituting our modified transverse wave vector in those models.

For the minimum self-healing distance we employ equation (6) and substitute the wave–number relations for $\tan \theta$ with and without the OAM contribution. For non-diffracting beams without OAM this comes to

$$z_{min,0} = \frac{a}{\tan \theta} = \frac{ak_{z,0}}{k_\rho} \qquad (12)$$

which is the relation according to the model developed in [8]. Extending this model with the contribution from OAM_$\ell$ to the transverse wave vector, we obtain

$$z_{min,l} = \frac{a}{\tan \theta} = \frac{ak_{z,l}}{\sqrt{k_\rho^2 + k_{T,l}^2}} \qquad (13)$$

where we note that the axial wave vector with OAM is now given by $k^2_{z,l} = k^2 - (k^2_\rho + k^2_{Tl})$. The radial and free-space wave vectors remain the same, so it becomes clear that the self-healing distance with OAM is always shorter than without OAM. Moreover, with our arrangement of using larger-aperture UCAs to source higher-order OAM beams and smaller ones for lower-order OAM so that all OAM beams have the *same beam-widths* (or cone angles, hence also equal beam radii, $R$) causes the OAM contribution to be greater *in proportion* to the order of OAM. That is because the magnitude of the tangential wave vector $|\vec{kT_1}|^2 = (\ell/R)^2$ increases with order $\ell$, but if all OAM modes originate from the same UCA aperture as in [11], then $R$ also increases proportionally so the tangential wave vector does not increase in magnitude with $\ell$. Therefore we can conclude that OAM-assisted self-healing leads to shorter minimum self-healing distances than those attained in beams without OAM, i.e. that

$$z_{min,l} < z_{min,0} \qquad (14)$$

The advantage of shorter self-healing distances past an obstacle with our equal-width OAM beams over those of non-OAM beams is summarized in Figure 13. The minimum self-healing distances for non-OAM beams and OAM beam up to order 64 are plotted against the distance of the obstruction from the transmitter, for a fixed link distance. The differences of the OAM-assisted self-healing minimum distances from that of a non-OAM beam are also plotted for clarity. Note that $R$ is assumed to increase with distance $z$ as it would in the far field past the non-diffracting region, with $\theta = 2.5$ degrees being our experimental half-cone angle of all our beams.

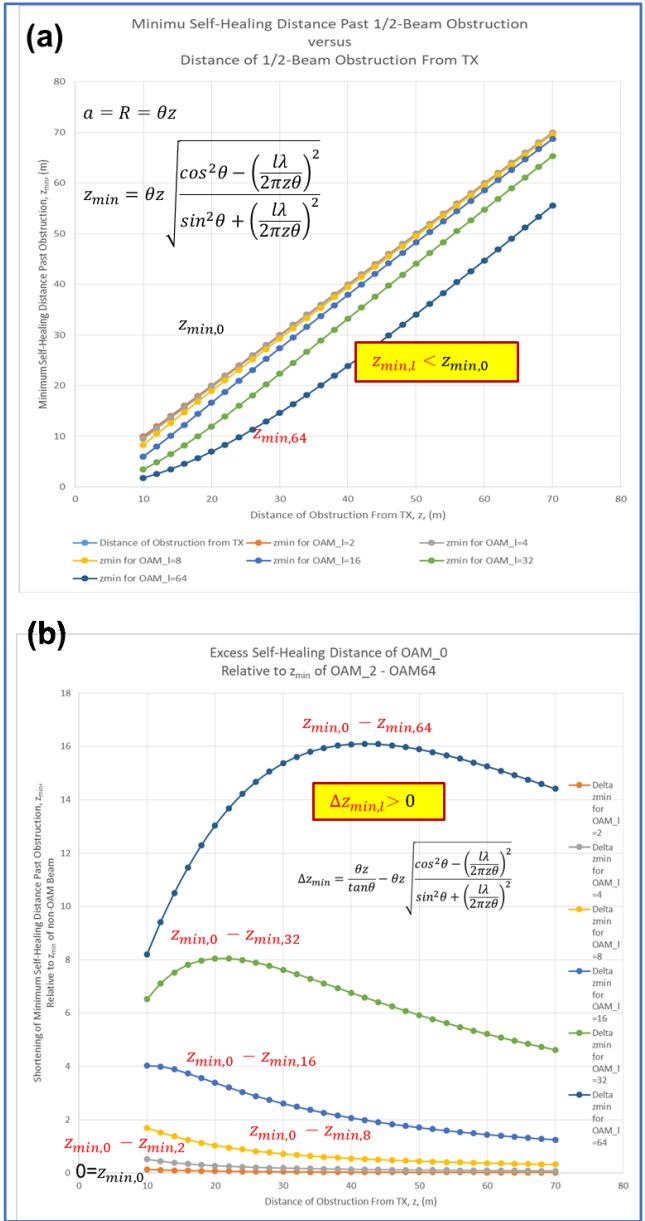

experiments, as it pertains to healing the entire beam at a given distance rather than only to the portion seen by our receiving antenna array. We chose the parameters according to our experimental setup again, beam waist corresponding to our beam-width increasing linearly with distance at half-cone angle $\theta = 2.5$ degrees, and obstruction width being half of the beam-width corresponding to our blockage of the right or left half of the OAM beam.

The resulting relations for the OAM-assisted model extension thus developed from [8] by extending their model to include the effect of our postulated azimuthal wave vector $k_{T\ell} = \ell/R$ are shown in Figure 14 alongside the plots. The plots show in dB the reciprocal of the self-healing "amount" defined in [8] for easier comparison with our obstruction-induced path losses, but again are not meant for a direct numerical comparison, since we were not quantifying the amount of self-healing relative to the entire beam as in [8], but only to the portion of the arc seen by the receiving array, which is much smaller. Hence our obstruction losses were much larger than those implied by the plots.

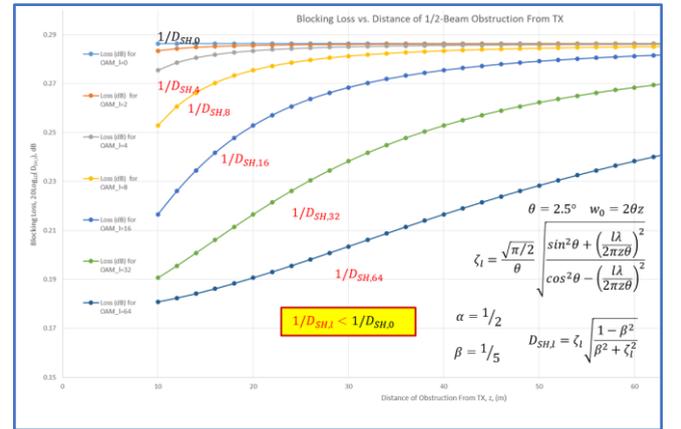

**Figure 14.** Relative obstruction-incurred losses in terms of reciprocal amounts of self-healing versus obstruction distance from transmitter, z, for OAM beams and non-OAM beam: Equal OAM angular beam-widths are assumed for the entire range of z. The features shown in red are due to assistance from OAM

**Figure 13.** Minimum self-healing distances versus obstruction distance from transmitter, z, for OAM beams and non-OAM beam: (a) Minimum self-healing distances, (b) difference of OAM-beam minimum distances from minimum distance for non-OAM beam. Equal OAM angular beam-widths are assumed for the entire range of z. The features shown in red are due to assistance from OAM.

The graphs are not meant as theoretical bases for comparison of our experimental results, but rather to show the relative improvements in $z_{min}$ resulting from extending theoretical models based on wave vector geometry and beam parameters to include the azimuthal transverse wave vector representing the assistance of OAM in self-healing. Since a small minimum self-healing distance is desired, it is evident that OAM improves the performance in this respect.

Another aspect of self-healing we compared with that of non-OAM beams was the "amount" of self-healing as defined in reference [8]. This is different from the differences in path losses incurred by insertion of the obstruction in our

Since smaller obstruction-incurred losses are desirable, it is again evident that the assistance from OAM improves the self-healing performance in this respect. Moreover, the improved self-healing performance is not restricted to the near-field DOF region as in [8], but it persists into the far field because the OAM phase factor in equation (2) is independent of propagation distance z.

What is also evident from Figures 13 and 14 is that the improvements expected from OAM-assistance in self-healing are most useful for high-order OAM beams, specifically much larger than the orders of OAM employed in our experimental investigations. While our OAM antennas is capable of supporting high-order OAM beams, the required DFT matrix feeds and probe assemblies would be impractically cumbersome to implement. To determine the highest-order OAM beam that can be radiated to the far field by a given size of circular array antenna, we assume a half-wavelength element spacing around its circumference. We also note that in order to propagate along it axis in the *z*-direction, the wave vector $k_z$ must have a real magnitude even beyond the DOF region. Accordingly, if we rearrange equation (9) to isolate it we obtain the axial wave vector as

$$k_z = \sqrt{k^2 - (k_\rho^2 + k_{T,l}^2)} \qquad (15)$$

Considering that the OAM beam must propagate beyond the DOF region into far field where the radial transverse wave vector is no longer effective, we require that $k^2 > k_{T,l}^2$. Recalling that $k=2\pi/\lambda$ and $k_{T\ell}=\ell/R$, we substitute them in (15) with $k_\rho \to 0$ and rearrange this condition as

$$\frac{2\pi}{\lambda} > \frac{l}{R} \qquad (16)$$

Finally we invoke the half-wavelength element spacing of *N* elements around the circumference $2\pi R$ so that $2\pi R=N\lambda/2$ and rearrange (16) to obtain the highest propagating OAM mode order as

$$l > N/2 \qquad (17)$$

In the case of our OAM antenna, the radius of each disc is large enough and the RF wavelength small enough that *N* is in the order of 100-200, so that beams up to OAM±64 characterized in Figures 13 and 14 could in principle be supported by it. By orienting the elements of the UCA radially and assuming circular polarization, the mutual-coupling matrix is circulant, so a DFT feed matrix effectively uncouples the OAM modes as explained in Appendix B.

## V. CONCLUSIONS AND FUTURE WORK

In this work we investigated the self-healing behavior of OAM beams of different orders and polarities to further characterize it in terms of the model developed in [1]. We conducted systematic and controlled experiments targeting the directionality, proportionality and persistence of the OAM self-healing mechanism with respect to the position of the beam obstruction. Our results were in every respect consistent with the hypothesized OAM-assisted self-healing mechanism, the essence of which was the addition of a tangentially-directed transverse wave vector to the radially-directed transverse wave vector and to the axial wave vector, all three being mutually orthogonal. We have thereby extended the geometric wave-vector based models of self-healing for non-diffracting beams in the literature to include OAM beams even beyond the non-diffracting region, into the far field. Although the various geometric aspects of our postulated OAM self-healing mechanism, such as the tilt angle of the helical Poynting vector, were identified in the literature, they were not incorporated into a self-healing mechanism, or the self-healing mechanisms that were developed for non-diffracting beams did not include the tangential transverse wave vector we formulated in our model.

We have therefore also developed a unified mechanism for OAM-assisted self-healing which includes non-diffracting beams in the depth-of-focus (near field) as well as OAM beams even in the far field region of propagation. We used this unified extended self-healing mechanism to explain why OAM beams have advantages over other, more conventional beams that do not carry OAM by adapting examples from the literature.

Furthermore, by treating the azimuthal phase factor of OAM beams as a product of our azimuthal transverse wave vector and the azimuthal propagation distance along the circumferential arc of the OAM beams, we account for all three directions of power flow in OAM beams in cylindrical coordinates and further justify our OAM-assisted geometric self-healing mechanism. This formulation also allows derivation of the orbital angular momentum from the tangential transverse wave vector, which is also included in the solution for the complex amplitudes of the fields of non-diffracting beams based on Maxwell's equations developed in Appendix A, providing a solid scientific foundation for our OAM-assisted self-healing model.

As for future work, we note that according to our OAM-assisted self-healing model, the benefit due to OAM increases with the azimuthal component of the transverse wave vector, $k_{T\ell}=\ell/R$. Accordingly we aim to increase its magnitude. There being only 2 parameters that define this wave vector, i.e. the OAM order $\ell$ and the beam radius *R*, we aim to optimize these to increase the OAM-assisted self-healing performance. That is, we aim to increase the OAM order while keeping their radii the same for all orders (to maintain the co-focusing property), and secondly to reduce the beam radius, *R*.

As mentioned in the previous section, increasing OAM order by using a larger DFT matrix feed to a UCA becomes impractical, even if the size of the UCA allows it to support such modes. A more practical approach would be to devise a kind of OAM-modal up-converter using a phase-gradient shaping dielectric prisms or lenses [14], or meta-surfaces as in [18].

Techniques for reducing the OAM beam radius would not only be helpful in enhancing their self-healing properties but also in improving the RX signal-to-noise ratios in OAM links. These could include lens [19] and sub-array techniques [20] applied to circular OAM antennas.

# VI. APPENDIX A: DERIVATION OF COMPLEX AMPLITUDES OF NON-DIFFRACTING OAM BEAMS IN CYLINDRICAL COORDINATES

In this Appendix we derive the complex amplitude distribution in the transverse plane of a non-diffracting OAM beam as particular solutions of the homogeneous wave equation, which in turn is derived from Maxwell's equations. It is intended as a more scientifically rigorous justification for the self-healing models and associated wave vectors, by showing how they emerge from the solution along with their relationships that were utilized in the body of this paper to formulate the OAM-assisted self-healing mechanism. While this treatment is not novel, it is deemed more accessible to the engineering community than the more general treatment found in the physics community as exemplified by [21]. It is instructive to note the assumptions and approximations made along the way in order to understand the applicability and limitations of this analysis in terms of the physical constraints.

We begin with the well-known Maxwell's equations for the electric and magnetic vector fields $E, H$:

$$\nabla \cdot E = \frac{\rho_e}{\epsilon_0} \tag{A1}$$

$$\nabla \cdot H = 0 \tag{A2}$$

$$\nabla \times E = -\mu_0 \frac{\partial H}{\partial t} \tag{A3}$$

$$\nabla \times H = \epsilon_0 \frac{\partial E}{\partial t} + J_e \tag{A4}$$

where $\rho_e$ is the electric charge density, $J_e$ is the vector electric current density and $\mu_0, \epsilon_0$ are the magnetic permeability and electrical permittivity of free space, respectively, with $\epsilon_0 \mu_0 = 1/c^2$, $c$ being the speed of light in free space. The first assumption we make is that we are interested in the propagating fields in the region away from the sources, where the current and charge densities are absent, i.e. $\rho_e=0$ and $J_e=0$. We note then that knowing only one of $E$ and $H$ fields enables one to find the other in equations (A3) and (A4). Accordingly, we proceed with equation (A3) and apply the curl operator $\nabla\times$ to both sides:

$$\nabla \times \nabla \times E = -\mu_0 \nabla \times \left(\frac{\partial H}{\partial t}\right) \tag{A5}$$

Inverting the order of the curl and time-derivative operations allows to substitute the left-hand side of (A4) for $\nabla\times H$ to yield an equation solely in terms of the electric field, $E$:

$$\nabla \times \nabla \times E = -\frac{1}{c^2}\left(\frac{\partial^2 E}{\partial t^2}\right) \tag{A6}$$

Next we substitute the identity $\nabla\times\nabla\times E = \nabla(\nabla \cdot E) - \nabla^2 E$ [22] for the left-hand side of (A6) noting that we are in the source-free region where $\nabla \cdot E=0$

$$\nabla^2 E - \frac{1}{c^2}\left(\frac{\partial^2 E}{\partial t^2}\right) = 0 \tag{A7}$$

We now invoke the restriction to **monochromatic** light (or narrow-band RF) at center radian frequency $\omega=2\pi f_{RF}$ and to **one-dimensional propagation along the z-axis** with propagation constant $k_z=2\pi/\lambda_z$, so the electric field is assumed to have the d'Alembert form, which we will analyze in Cartesian coordinates to begin with:

$$E(x,y,z,t) = E(x,y,z)e^{j(\omega t - k_z z)} \tag{A8}$$

Upon substitution in (A7), we obtain the following:

$$\nabla^2\left(E(x,y,z)e^{j(\omega t - k_z z)}\right) - \frac{1}{c^2}\left(\frac{\partial^2 E(x,y,z)e^{j(\omega t - k_z z)}}{\partial t^2}\right) =$$
$$= e^{j\omega t}\nabla^2(E(x,y,z)e^{-jk_z z}) - \frac{e^{-jk_z z}}{c^2}E(x,y,z)\frac{\partial^2(e^{j\omega t})}{\partial t^2} = 0 \tag{A9}$$

noting that $(x, y, z)$ are independent of $t$ and each other. We continue expanding (A9) to yield

$$e^{j(\omega t)}\nabla^2(E(x,y,z)e^{-jk_z z}) - \frac{e^{-jk_z z}}{c^2}E(x,y,z)\frac{\partial^2(e^{j\omega t})}{\partial t^2} = 0 \tag{A10}$$

We note that $\frac{\partial^2(e^{j\omega t})}{c^2 \partial t^2} = -e^{j\omega t}(\omega/c)^2 = -e^{j\omega t}k^2$ where $k=2\pi/\lambda$ is a hard physical constant for a given RF wavelength, $\lambda$, so (A10) becomes

$$e^{j(\omega t)}\nabla^2(E(x,y,z)e^{-jk_z z}) + e^{j(\omega t - k_z z)}k^2 E(x,y,z) = 0 \tag{A11}$$

Expanding the Laplacian spatial derivatives and re-arranging some common factors produces

$$e^{j(\omega t - k_z z)}\left(\frac{\partial^2 E(x,y,z)}{\partial x^2} + \frac{\partial^2 E(x,y,z)}{\partial y^2}\right) + \cdots$$
$$\cdots + e^{j\omega t}\frac{\partial^2(E(x,y,z)e^{-jk_z z})}{\partial z^2} + \cdots$$
$$\cdots + e^{j(\omega t - k_z z)}k^2 E(x,y,z) = 0 \tag{A12}$$

It is instructive to expand the derivative with respect to z and expose the subsequent approximations for the field $E(x, y, z)$.

For simplicity we will ignore the common exponential factor $e^{j(\omega t - k_z z)}$ and rewrite (A12) as

$$\left(\frac{\partial^2 E(x,y,z)}{\partial x^2}+\frac{\partial^2 E(x,y,z)}{\partial y^2}\right)+\cdots$$
$$\cdots + e^{jk_z z}\left(\frac{\partial^2 (E(x,y,z)e^{-jk_z z})}{\partial z^2}\right)+\cdots$$
$$\cdots + k^2 E(x,y,z) = 0 \qquad (A13)$$

Now we expand the second bracketed term on the left-hand side as

$$\left(\frac{\partial^2 (E(x,y,z)e^{-jk_z z})}{\partial z^2}\right)=\frac{\partial}{\partial z}\left(\frac{\partial (E(x,y,z)e^{-jk_z z})}{\partial z}\right)=\cdots$$
$$\cdots = \frac{\partial}{\partial z}\left(\frac{E(x,y,z)\partial e^{-jk_z z}}{\partial z}+\frac{e^{-jk_z z}\partial E(x,y,z)}{\partial z}\right)$$
$$(A14)$$

Continuing with the right-hand side of (A14) we expand it further as

$$\left(\frac{\partial^2 (E(x,y,z)e^{-jk_z z})}{\partial z^2}\right)=\cdots$$
$$\cdots = \frac{\partial}{\partial z}\left(\frac{E(x,y,z)\partial e^{-jk_z z}}{\partial z}+\frac{e^{-jk_z z}\partial E(x,y,z)}{\partial z}\right)=\cdots$$
$$=\frac{\partial}{\partial z}\left(\frac{E(x,y,z)\partial e^{-jk_z z}}{\partial z}\right)+\frac{\partial}{\partial z}\left(\frac{e^{-jk_z z}\partial E(x,y,z)}{\partial z}\right)=\cdots$$
$$\cdots = \left(E(x,y,z)\frac{\partial^2 (e^{-jk_z z})}{\partial z^2}+\frac{\partial e^{-jk_z z}}{\partial z}\frac{\partial E(x,y,z)}{\partial z}\right)+\cdots$$
$$\cdots + \left(e^{-jk_z z}\frac{\partial^2 E(x,y,z)}{\partial z^2}+\frac{\partial E(x,y,z)}{\partial z}\frac{\partial e^{-jk_z z}}{\partial z}\right)$$
$$(A15)$$

The right-hand side of (A15) simplifies with $\frac{\partial e^{-jk_z z}}{\partial z}=-jk_z e^{-jk_z z}$ and $\frac{\partial^2 (e^{-jk_z z})}{\partial z^2}=-k_z^2 e^{-jk_z z}$, so that (A15) becomes

$$\left(\frac{\partial^2 (E(x,y,z)e^{-jk_z z})}{\partial z^2}\right)=\cdots$$
$$\cdots = \left(-k_z^2 e^{-jk_z z}E(x,y,z)-jk_z e^{-jk_z z}\frac{\partial E(x,y,z)}{\partial z}\right)+\cdots$$
$$\cdots + \left(e^{-jk_z z}\frac{\partial^2 E(x,y,z)}{\partial z^2}-jk_z e^{-jk_z z}\frac{\partial E(x,y,z)}{\partial z}\right)$$
$$(A16)$$

which further simplifies to

$$\left(\frac{\partial^2 (E(x,y,z)e^{-jk_z z})}{\partial z^2}\right)=\cdots$$
$$\cdots = e^{-jk_z z}\left(\begin{array}{c}-k_z^2 E(x,y,z)-\cdots\\ \cdots - j2k_z\frac{\partial E(x,y,z)}{\partial z}+\frac{\partial^2 E(x,y,z)}{\partial z^2}\end{array}\right)$$
$$(A17)$$

We are now ready to substitute the right-hand side of (A17) back into (A13). Noting immediately that the exponential factors cancel, we obtain

$$\left(\frac{\partial^2 E(x,y,z)}{\partial x^2}+\frac{\partial^2 E(x,y,z)}{\partial y^2}\right)-k_z^2 E(x,y,z)-\cdots$$
$$\cdots - j2k_z\frac{\partial E(x,y,z)}{\partial z}+\frac{\partial^2 E(x,y,z)}{\partial z^2}+k^2 E(x,y,z)=0$$
$$(A18)$$

With some foresight we define $k_\rho^2 = k^2 - k_z^2$ rewrite (A18) as

$$\left(\frac{\partial^2 E(x,y,z)}{\partial x^2}+\frac{\partial^2 E(x,y,z)}{\partial y^2}+\frac{\partial^2 E(x,y,z)}{\partial z^2}\right)+\cdots$$
$$\cdots + k_\rho^2 E(x,y,z) - j2k_z\frac{\partial E(x,y,z)}{\partial z}=0$$
$$(A19)$$

The first approximation we make is that the monochromatic wave propagating in the *z*-direction is confined to small angles *θ* around the *z*-axis, which is termed the **paraxial approximation** whereby $\frac{\partial^2 E(x,y,z)}{\partial z^2}\ll 2k_z\frac{\partial E(x,y,z)}{\partial z}$, or equivalently, $\frac{\partial E(x,y,z)}{\partial z}\ll 2k_z = 2\pi/\lambda_z$. This is common in optics where the wavelengths are very small compared to the dimensions of the optical components and propagation distances, even in the near-field. It essentially means the field amplitude varies very slowly with propagation distance along *z* compared to its propagation constant for the *z*-direction $k_z$. The paraxial propagating field then obeys

$$\left(\frac{\partial^2 E(x,y,z)}{\partial x^2}+\frac{\partial^2 E(x,y,z)}{\partial y^2}\right)+k_\rho^2 E(x,y,z)-\cdots$$
$$\cdots - j2k_z\frac{\partial E(x,y,z)}{\partial z}=0$$
$$(A20)$$

Now we impose the condition that our paraxial wave is **non-diffracting**, whereby $\frac{\partial E(x,y,z)}{\partial z}\approx 0$ so *E(x, y, z)* essentially becomes *E(x, y)*, meaning that its profile in the transverse *(x, y)* plane remains unchanged with propagation distance in the *z* direction. What remains of (A20) is then

$$\left(\frac{\partial^2 E(x,y)}{\partial x^2}+\frac{\partial^2 E(x,y)}{\partial y^2}\right)+k_\rho^2 E(x,y)=0$$
$$(A21)$$

In relation to self-healing, it appears that the same mechanism that is responsible for the non-diffracting property we demand of *E(x, y)* is also responsible for its self-reconstruction property. Normally a beam propagating from a given finite aperture will expand to be wider than the aperture as it propagates, and will form a diffraction pattern of the aperture in the far field as is well known from the Fourier transform relation between aperture distribution and the far-field beam shape or diffraction pattern. Here we place a condition on the aperture field *E(x, y)* to keep that from happening, implying that some mechanism will be opposing the beam's tendency to expand, i.e. opposing its diffraction and maintaining its transverse structure. This tendency to maintain structure will also manifest as a tendency to re-form the structure when it is disturbed, ***hence the non-diffracting property leads to self-healing***.

With some foresight, we note the reference to a finite aperture as the origin of the non-diffracting wave, so we expect the propagation distance where it maintains its non-diffracting property to also be a finite maximum value $z_{max}$ as indicated in Figure 10 in the main body of this paper. It will emerge in the end that the solution for *E(x, y)* is infinite in the transverse plane so there is no limit to the non-diffracting distance except for the practical limit we must impose on the physical extent of *E(x,y)* after the theoretical solution is obtained. This is the depth of focus (DoF) region in the near field of the aperture, where self-healing occurs even without the beam possessing orbital angular momentum (OAM).

Before we impose the additional condition of OAM on *E(x,y)*, we formulate the right-angle triangle relation according to Pythagoras among the wave-numbers or propagation constants that we have defined so far without orbital angular momentum, or OAM_0, as in Figure A1.

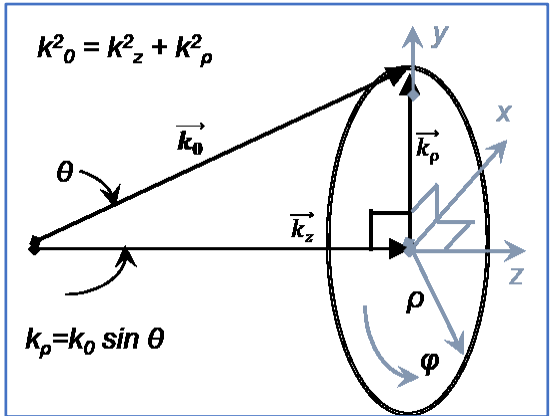

**Figure A1. Relation among wave vectors in non-diffracting beam propagating in z direction, with free-space propagation constant (without OAM) being $k=k_0=2\pi/\lambda$, and cylindrical transverse coordinates *(ρ, φ)*.**

Now we change to cylindrical coordinates in anticipation of the azimuthally-varying phase of *E(x, y)* due to the ***orbital angular momentum (OAM) that we now impose*** on it as another condition. This transformation of coordinates will facilitate the solution as that for *E(ρ, φ )*. Its relation to self-healing will emerge from the solution for *E(ρ, φ )* and the associated wave-number and momentum relations. The cylindrical coordinates *(ρ, φ, z)* are related to the Cartesian ones as $x = \rho \cos\varphi, y = \rho \sin\varphi, z = z$. Note that there is no change in *z*, and the same exponential factor $e^{j(\omega t - k_z z)}$ applies to the complete solution of (A21). Using [10] and [22] as guides, we proceed as follows:

We note that (A21) contains the Laplacian operator $\nabla^2 = \frac{\partial^2}{\partial x^2} + \frac{\partial^2}{\partial y^2} + \frac{\partial^2}{\partial z^2}$ with $\frac{\partial^2}{\partial z^2} = 0$ so we transform that into cylindrical coordinates. Omitting the tedious details of computing and applying the scaling factors according to the formalism of orthogonal curvilinear coordinates [22], we obtain the transformed version of (A21) as

$$\frac{1}{\rho}\frac{\partial}{\partial \rho}\left(\rho \frac{\partial E(\rho,\varphi)}{\partial \rho}\right) + \frac{1}{\rho^2}\frac{\partial^2 E(\rho,\varphi)}{\partial \varphi^2} + k_\rho^2 E(\rho,\varphi) = 0$$
(A22)

Note that the time- and *z*-dependent exponential factor still applies to the complete solution for *E(ρ, φ)* as before. Next we assume that the solution can be expressed in the form of separation of variables [10] as

$$E(\rho,\varphi) = A(\rho)\Psi(\varphi)$$
(A23)

and we further assume that $\Psi(\varphi)$ has the form

$$\Psi(\varphi) = e^{jl\varphi}$$
(A24)

with *ℓ* being any integer, because we want the phase to be continuous and periodic in *φ* . (This also corresponds to the phase-continuity condition imposed in [6].) Substitution of (A24) in (A23) and subsequently into (A22) finally produces

$$e^{jl\varphi}\left(\frac{\partial^2 A(\rho)}{\partial \rho^2} + \frac{1}{\rho}\frac{\partial A(\rho)}{\partial \rho} + \left(k_\rho^2 - \frac{l^2}{\rho^2}\right)A(\rho)\right) = 0$$
(A25)

where $k_\rho = k_0 \sin\theta = \sqrt{k_0^2 - k_z^2}$ is identified in Figure A1. The *ρ* –dependent factor is in the form of Bessel's equation [10] which has solutions in the form of Bessel functions of order *ℓ*, i.e. the radial factor of the amplitude of the E-field in the aperture has the form $A(\rho) = \mu J_l(k_\rho \rho)$ so that the complete solution is

$$E(\rho,\varphi,z,t) = \mu e^{j(\omega t - k_z z)} e^{jl\varphi} J_l\left(\frac{2\pi}{\lambda}\rho \sin\theta\right)$$
(A26)

where we used $k_\rho = k_0 \sin\theta = \frac{2\pi}{\lambda}\sin\theta$ . Note that this agrees with equation (3) in the main body of this paper, which corresponds to the far field radiation pattern of the aperture and is also an approximation. Because the Bessel function extends infinitely in *ρ*, we limit it to a finite extent, thus obtaining a finite depth of focus in *z* also.

We observe that equation (A25) contains the scalar term $\left(k_\rho^2 - \frac{l^2}{\rho^2}\right)$ where we identify our friend from [6] as

$$k_{T,l} = \left(\frac{l}{\rho}\right), \qquad (A27)$$

so it is a legitimate wave vector in the tangential $\varphi$-direction as depicted in Figure 11 in the body of the paper. It is entirely due to the OAM of the non-diffracting wave and plays an integral role in the solution of the radial field equation (A25). It is not dependent on $z$ so it persists in the solution for all values of z even beyond the non-diffracting region dictated by a finite extent of $\rho = R$, where it becomes

$$k_{T,l} = \left(\frac{l}{R}\right), \qquad (A28)$$

exactly as envisioned in [6].

The role of the *k*-vectors is important in establishing the momentum carried by the wave. To illustrate loosely, recall that in mechanics the momentum *p* of a mass *m* moving with velocity *v* is *p=mv*. It is also the derivative of its kinetic energy *ε* with respect to velocity as

$$p = \frac{d\varepsilon}{dv} = \frac{d(mv^2/2)}{dv}, \qquad (A29)$$

However, photons do not have mass when propagating with velocity *v=c*, but do have energy given by the Einstein relation for the photo-electric effect *ε=hf*, where *h* is Planck's constant and *f=c/λ* is the photon's frequency and *λ* its wavelength. We can therefore write its momentum as

$$p = \frac{d\varepsilon}{dc} = \frac{d(hf)}{dc} = h\frac{d(c/\lambda)}{dc} = \frac{h}{\lambda}, \qquad (A30)$$

This can now be written in terms of the wave number *k=2π/λ* in the direction of the k-vector in Figure 12, as

$$p = \frac{hk}{2\pi} = \bar{h}k \qquad (A31)$$

where $\bar{h} = \frac{h}{2\pi}$ is the reduced Planck's constant. We can apply the same relation as (A31) to our transversal wave vector to find the corresponding orbital angular momentum, OAM, but we must cross-multiply it by its moment arm, i.e. the radius $\rho$. Designating the OAM by *q*, we can write

$$q = \frac{h}{2\pi}\left(\frac{l}{\rho}\right) \times \rho = \bar{h}l \qquad (A32)$$

These relations also agree with reference [21], which lends legitimacy to our concept of treating the quantity *ℓ/R* as a tangential wave vector in the $\varphi$-direction, thus ***unifying*** it with the other wave vectors in the ***theory of self-healing OAM beams.***

## VI. APPENDIX B: DECOUPLING THE OAM MODES OF A UNIFORM CIRCULAR ARRAY OF RADIALLY-ORIENTED CIRCULARLY-POLARIZED ELEMENTS USING A DFT FEED MATRIX

In this Appendix we address mutual coupling among elements of a uniform circular array, and also among a UCA of probes feeding identical radial sectors of such an array.

When the radiating elements of a UCA are oriented along regularly-spaced radii, each sees the same electromagnetic environment and couples to it the same way. This makes the mutual coupling matrix, $\mathbf{M}$, have a circulant structure, whereby each row (or column) is a circular shift of the previous row (or column). The currents induced in each radiating element due to mutual coupling with its neighbours add to its excitation current to form the total current responsible for its radiated fields as well as its reactive near fields.

It can be proven that any circulant matrix is diagonalized by a Discrete Fourier Transform (DFT) matrix of the same square dimensions[1]. We apply this fact to show how a DFT feed matrix effectively collapses the mutual coupling among the elements of a UCA when viewed from its inputs, which correspond to excitation ports of Orbital Angular Momentum (OAM) modes, a.k.a. phase modes. This property greatly facilitates generating OAM modes with robust purity and orthogonality in the circumferential (azimuth) direction. It is important to note that the requisite polar symmetry of the UCA elements includes their polarization, which translates to a requirement for circular polarization. Any polarization desired for the radiated fields can be synthesized from Left-Handed and Right-Handed Circular polarizations (LHCP and RHCP, respectively), but a separate DFT matrix feed is required for each polarization [23].

Because the UCA and its feed structure is ideally a lossless passive structure, reciprocity applies. Accordingly, we conduct the ensuing development in the receiving mode, knowing that it applies equally in the transmitting mode of the UCA.

Consider a UCA equipped with a DFT matrix feed network, receiving a waveform corresponding to an n-th order OAM mode that excites currents on the UCA elements represented by elements of vector $\mathbf{R}$. In the absence of mutual coupling and other imperfections, the entire signal of that waveform would appear at the n-th feed port of the DFT matrix, with the other ports showing zero signals, as denoted by vector $\mathbf{P}$.

$$\mathbf{P} = \begin{bmatrix} 0 \\ \vdots \\ p_n \\ \vdots \\ 0 \end{bmatrix} = [DFT]\mathbf{R} \qquad (B1)$$

With additional currents induced through mutual coupling among the UCA elements the received signal vector acquires additional components and is denoted by $\mathbf{P'}$ as

$$\mathbf{P'} = [DFT][I + M]\mathbf{R} = \mathbf{P} + [DFT]\mathbf{MR} \qquad (B2)$$

Recall that matrix $\mathbf{M}$ is circulant, so it is diagonalized by the DFT matrix to produce a diagonal matrix $\mathbf{C}$ according to

$$\mathbf{M} = [DFT]^{-1}\mathbf{C}[DFT] \qquad (B3)$$

Substituting this in (B2) leads to the resultant received signal vector $\mathbf{P'}$ in terms of the ideal vector $\mathbf{P}$, but containing effects of mutual coupling as

$$\mathbf{P'} = \mathbf{P} + [DFT]\mathbf{MR} = \mathbf{P} + \cdots \\ \cdots + [DFT][DFT]^{-1}\mathbf{C}[DFT]\mathbf{R} \qquad (B4)$$

which simplifies to

$$\mathbf{P'} = \mathbf{P} + \mathbf{CP} \qquad (B5)$$

If the n-th diagonal element of diagonal matrix $\mathbf{C}$ is $c_{nn}$, then the entire effect of mutual coupling currents manifests only on the n-th element of $\mathbf{P'}$, which amounts to scaling it by $(1+c_{nn})$ and not affecting any other received signal elements of $\mathbf{P}$. Therefore one can write

$$\mathbf{P'} = \begin{bmatrix} 0 \\ \vdots \\ p'_n \\ \vdots \\ 0 \end{bmatrix} = \begin{bmatrix} 0 \\ \vdots \\ p_n \\ \vdots \\ 0 \end{bmatrix} + c_{nn} \begin{bmatrix} 0 \\ \vdots \\ p_n \\ \vdots \\ 0 \end{bmatrix} \qquad (B6)$$

which shows that none of the UCA elements' signals from the ideal received OAM mode couple to the desired received n-th OAM mode, but merely scale it by a generally complex constant that is easy to compensate.

---

[1] See for example here:
https://web.mit.edu/18.06/www/Spring17/Circulant-Matrices.pdf

**MAREK KLEMES** (Life SM-IEEE) was born in Brno, Czech Republic. 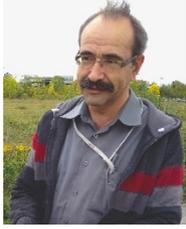 He obtained his B.A.Sc. (hons) and M.A.Sc. in Electrical Engineering from the University of Toronto (Toronto, Canada) in 1981 and 1983, respectively, and Ph.D. from Carleton University (Ottawa, Canada) in Electronics in 1997. From 1983 to 1996, he was adaptive systems engineer at Canadian Marconi Company where he developed adaptive antenna array subsystems for UHF tactical radios for US Army (CECOM) and commercial wireless base-stations, and super-resolution direction-finding test bench for Canadian department of defense (DREO). He then worked as a senior radio systems engineer at Nortel (Ottawa, 1996 - 2001), and Dragonwave Inc. (Kanata, 2001-2012), involved in propagation channel modelling and link budget design, lightning protection and radio system design and testing, and detailed simulation, design, integration and prototyping of modem subsystems such as PLLs, adaptive equalizers, AGCs, cross-polarization interference canceller (XPIC), diversity combiners and beam-formers. Since 2012 he has been a senior wireless research engineer at Huawei Canada R&D Center (Kanata), working on 5G and 6G millimeter-wave waveforms and associated RF analog and adaptive subsystems and signal-processing algorithms. His present area of interest is in the spatial multiplexing using Orbital Angular Momentum (OAM) waves, OAM antenna design and propagation phenomena, MIMO reception techniques, blind signal-separation (BSS) adaptive algorithms and wireless and optical sensing techniques. He holds 10 patents, has published 20 papers and enjoys all aspects of rock-hounding and antique radio experimentation.

**LAN HU** received the M.Sc degree from the Shanghai Jiaotong University, China, and the Ph.D degree from the University of Limerick, Ireland. From 1996 to 2009, she worked in Nortel wireless advanced technologies division. She joined Huawei Ottawa advanced wireless technologies Lab in 2009. As a senior principal researcher in wireless technologies, she has involved many projects from 2G to 6G research and product development.

**GREGORY J. BOWLES** received his BEng in Electrical Engineering (Communications) from the Canadian Royal Military College (RMC) in 1995 with focus on computer aided design (CAD), electromagnetics and circular polarization. He joined Nortel Networks in 1998 where he specialized in improved corporate design techniques, CAD work flow and high efficiency systems including novel Doherty amplifier architectures. In 2009, he joined the Ottawa Advanced Wireless Technology Lab of Huawei Technologies Canada Co., Ltd with an initial focus on high efficiency power amplifier designs before migrating to transceiver and Butler Matrix designs with a particular concentration on how they apply to Orbital Angular Momentum (OAM) based systems.

**ALIREZA GHAYEKHLOO** received the B.Sc. degree in electrical engineering from Mazandaran University (Babol Noshirvani University of Technology (NIT)), Babol, Iran, the M.Sc. degree in electrical engineering from the Iran University of Science and Technology, Tehran, Iran, and the Ph.D. degree in electrical engineering from Semnan University, Semnan, Iran, in 2012, 2014, and 2020, respectively. From 2015 to 2018, he was a Lecturer providing engineering and physics knowledge to younger generations and helping enhance the curriculum. During 2018–2019, he was with Center Energy, Materials and Telecommunications, Institut National de la Recherche Scientifique, Montreal, QC, Canada, on advanced electromagnetic structures to control wave scattering and propagation. He was a PostDoc Fellow with the Université du Québec en Outaouais, Gatineau, QC, Canada for two years, and currently is an associated researcher at Huawei Canada R&D Center (Kanata) working on societal and industry-driven proposals for next generation wireless technology. His research interests include advanced electromagnetics structures, antennas and microwave engineering, complex media, electromagnetics scattering, and multiphysics.

**MOHAMMAD AKBARI** (Member, IEEE) received the Ph.D. degree in electrical and computer engineering from Concordia University, Montreal, QC, Canada, in 2018. 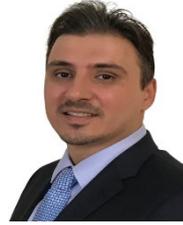 He was a postdoctoral fellow with the University of Waterloo, Waterloo, ON, Canada, in 2019. Also, in 2020, he was a member of the Electromagnetic Vision (EMVi) Research Laboratory, McMaster University, Hamilton, ON, Canada. He joined Huawei Technologies Canada Co. Ltd., Canada Research Centre, Kanata, in 2021 as a senior engineer. He is the author or coauthor of more than 80 peer-reviewed scientific journal articles and international conference papers. His current research interests include reconfigurable antennas, phased array antennas, metasurfaces and metamaterials, miniaturisation approaches, mutual coupling reduction techniques, feeding networks, RCS reduction, polarizers, and ultra-wideband (UWB) technology. He was a recipient of the Postdoctoral Fellowships from the Fonds de Recherche du Québec—Nature et Technologies (FRQNT) in 2019, the PERSWADE/NSERC-Create Training Programme Award in 2017, the Accelerator Award in 2017, the Graduate Concordia Merit Scholarship in 2016, and the Concordia University International Tuition Fee Remission in 2014. He served as a reviewer for the IEEE Antennas and Propagation Society's journals and magazines.

**SOULIDETH THIRAKOUNE** received the BASc degree in Electrical Engineering from the University of Ottawa, Ottawa, ON, Canada, in 1999. From 1999 to 2011, he carried out research at the Communication Research Centre Canada, Ottawa, as a Microwave Antenna Designer on printed antennas, dielectric-resonator antennas and holographic antennas. He joined Huawei Technologies Canada Co. Ltd, Canada Research Center, Kanata, in 2012 as a mmWave Antenna Design Engineer. He is working on various projects involving a wide range of antennas and arrays, including novel printed antennas, multilayer microstrip antennas, microwave lenses, active integrated antennas, and Waveguide antennas.

**MICHAEL SCHWARTZMAN** received his MSc Mechanical Engineer with experience in machine development and design, electronic packaging design, finite element analysis and structural strength calculations, from conceptual design and prototyping to detailed design and manufacturing.

**KEVIN ZHANG** received Master Degree and Bachelor Degree of EE from BUAA. Main focus on signal processing, hardware and FPGA design. Have worked in multiple telecom companies as senior engineering position.

**TAN HUY HO** received the BS degree in EE from the University of Victoria in 1996 and a MS degree from Carleton University in 2014. He has been in the wireless and datacom industries for over 28 years working in companies including start-ups, as well as, large corporations such as Nortel, Ciena Corporation, Spirent Communications, Ericsson and currently Huawei Technologies Canada since 2014. His current focus and interest at Huawei is research in mmW/sub-THz wireless system architecture and technologies, particularly, microwave-photonics technologies for 6G wireless integrated sensing and communications.

**DAVID WESSEL** has been Technical VP at Huawei Technologies Canada. Employed at Huawei since 2009. Previously with Nortel from 1991 to 2009 and Microwave Modules and Devices from 1987 to 1991. He is a graduate of the University of Wisconsin–Madison.

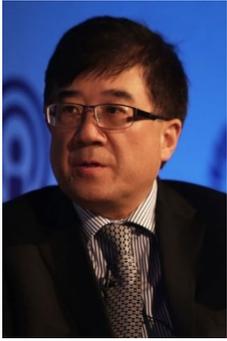

**Dr. WEN TONG** is the CTO, Huawei Wireless. He is the head of Huawei wireless research. In 2011, Dr. Tong was appointed the Head of Communications Technologies Labs of Huawei, currently, he is the Huawei 5G chief scientist and leads Huawei's 10-year-long 5G wireless technologies research and development.

Prior to joining Huawei in 2009, Dr. Tong was the Nortel Fellow and head of the Network Technology Labs at Nortel. He joined the Wireless Technology Labs at Bell Northern Research in 1995 in Canada.

Dr. Tong is the industry recognized leader in invention of advanced wireless technologies, Dr. Tong was elected as a Huawei Fellow and an IEEE Fellow. He was the recipient of IEEE Communications Society Industry Innovation Award for "the leadership and contributions in development of 3G and 4G wireless systems" in 2014, and IEEE Communications Society Distinguished Industry Leader Award for "pioneering technical contributions and leadership in the mobile communications industry and innovation in 5G mobile communications technology" in 2018. He is also the recipient of R.A. Fessenden Medal. For the past three decades, he had pioneered fundamental technologies from 1G to 6G wireless and Wi-Fi with more than 520 awarded US patents.

Dr. Tong is a Fellow of Canadian Academy of Engineering, Fellow of Royal Society of Canada and he serves as Board of Director of Wi-Fi Alliance.